\documentclass[doublecol,figures]{epl2} 
\def\aa{\AA\,} 

\newcommand{\bea}{\begin{eqnarray}} 
\newcommand{\eea}{\end{eqnarray}}
\def\APL{ Appl. Phys. Lett.}

\def\JCP{ J. Chem. Phys.}

\def\PRL{ Phys. Rev. Lett.} 
 
\def\PRE{{ Phys. Rev.} E}

\def\RMP{ Rev. Mod. Phys.}

\usepackage{graphicx} 
\begin{document}



\title{Charge transport in bacteriorhodopsin monolayers: The contribution of conformational change
to current-voltage characteristics}
\shorttitle{bR: Conformational change and charge transport in bacteriorhodopsin} 

\author{E. Alfinito\inst{1,2} \and L. Reggiani\inst{1,2}}
\shortauthor{E. Alfinito \etal}

\institute{                    
  \inst{1} Dipartimento di Ingegneria dell'Innovazione, Universit\`a del Salento - via Arnesano, I-73100 Lecce, Italy, EU\\
  \inst{2} CNISM - Consorzio Nazionale Interuniversitario per le Scienze Fisiche della Materia - via della Vasca Navale 84,I-00146 Roma, Italy, EU.
}

\abstract{
When moving from native to light activated bacteriorhodopsin, modification of charge transport consisting of an increase of conductance is correlated to the protein conformational change. 
A theoretical model based on a map of the protein tertiary structure into a resistor network 
is implemented to account for a sequential tunneling mechanism of charge transfer through neighbouring amino acids.
The model is validated by comparison with current-voltage experiments. 
The predictability of the model is further tested on bovine rhodopsin, a
G-protein coupled receptor (GPCR) also sensitive to light. In this case, results show an opposite behaviour with a decrease of conductance
in the presence of light.}

\maketitle

\section{Introduction}
In recent years, several papers reported on the electrical properties
of single proteins inserted into hybrid systems
\cite{Rinaldi,Bizzarri,Jin,Jin1,Gomila}.
The principal aim of these researches is to explore the possibility to design
nanodevices, mainly nanobiosensors, with extreme sensibility and specificity.
Most work was devoted to metalloproteins, which exhibit semiconductor-like
conductivities, but increasing attention has been also addressed to bacteriorhodopsin
(bR), the light sensitive protein present in \textit{Archaea}. 
Bacteriorhodopsin exhibits a conductivity close to that of an insulator, and is robust against thermal,
chemical and photochemical degradation \cite{Jin, Jin1}. 
It is a protein (opsin)-chromophore complex that
underlies photon-activated modifications in which the chromophore
changes its structure from the \textit{all-trans} to the \textit{13-cis}.  
The chromophore evolution induces multiple protein
transformations, which fastly drive the protein from the K$_{590}$  to the L$_{550}$ state. 
In the slow transition from this state to the M$_{410}$, a proton is released. 
Junctions prepared with bR monolayers contacted with Au electrodes,
show non-linear \textit{I-V} characteristics both in the dark (or blue light) and in the presence
of green light \cite{Jin, Jin1, Gomila}. It is generally assumed that green light produces a more intense current
as a consequence of the modification of the chromophore-opsin complex. 
Although the experimental results are not completely assessed from the quantitative side \cite{Jin1}, 
the main results of available experiments can be summarized as follows.
(i) A conductivity about four orders of magnitude higher than that of a homogeneous layer of similar 
dielectric material was observed. 
(ii) The current voltage (\textit{I-V}) characteristics exhibit a super-Ohmic behaviour pointing to a tunnelling 
charge-transfer mechanism.
(iii) The presence of green light significantly increases the electrical conductivity up to about a factor of three with respect to the dark value.
(iv) The \textit{I-V} 
sensitivity to the green light is washed out by substituting the original chromophore with one that is not sensible to light.
(v) When the chromophore of the protein is completely removed, the conductivity is suppressed by three orders of magnitude lower, taking the value of a standard
dielectric. 
Despite the presence of these experiments, the existing theoretical approaches only provide a phenomenological interpretation of some \textit{I-V} characteristics in terms of metal-insulator-metal tunneling theories \cite{Gomila,Reed}.
\par
The aim of this work is to develop a theoretical model able to capture the main features of the above experiments
and sufficiently general to constitute a new framework for describing electron transport properties through proteins. To this purpose the protein conductivity is modelled by a sequential tunneling  mechanism through
the protein amino acids, whose positions pertain to a specific conformation. Thus, the conductivity is directly connected to the protein tertiary structure
and a light induced structural change will result in conductivity change which can be quantitatively determined.
\section{The network tunneling model}
As general features,
the protein electrical conductivity is close to that of an insulator, but it can increase significantly in particular environmental conditions, like in organic semiconductors \cite{Baranovski}.
Furthermore, the microscopic mechanism of charge transport in proteins is very articulated  and includes electron, ion, and proton transport \cite{Warshel} in a yet not completely understood sequence. 
The difficulty to discriminate among these kinds of transport is further enhanced by the lack of experiments, mainly related to the difficulty of working with biological materials at the nanoscale.
The most qualified mechanisms to describe electron transport  in proteins are the hopping/tunneling processes  \cite{Frauenfelder, Canters}.
In this framework, the transport model which assumes the existence of privileged pathways of sequential tunneling  has gained a wide consensus \cite{Onuchic}.    By embracing this picture, we describe the protein as a network 
and the electron transport as due to a sequential tunneling along preferential channels.
In doing so, we produce a map of the protein by taking  
from the protein data base (PDB) \cite{PDB} its tertiary structure, and setting
up a graph whose nodes correspond to the amino acids (identified by the $\alpha$-carbons) \cite{Tirion,Nano}. Two nodes are then connected with a link only if they can interact within a given 
interaction radius $R_c$. 
\par
At this stage, we neglect thermal fluctuations because the relevant signals of experiments do not suggest their evidence.
\par
To the purpose of investigating the conduction properties of the protein,
the graph is turned into a resistor network:
Each link is replaced by an elemental resistance, which mimics the different
charge transfer property between neighbouring amino acids associated with their mutual distance. 
To detect the differences in conduction properties between two
protein states, as those produced by the conformational change, we adopt the simplest parameter model by 
choosing the same resistivity for each amino acids.
In this way, the tertiary structure and the interaction radius are the relevant input parameters of the theory. 
\par 
As elemental resistance of the network we take 
\begin{equation}
R_{i,j}=\frac{l_{i,j}}{{\mathcal{A}}_{i,j}}\rho
\label{1}
\end{equation}   
where 
${\mathcal{A}}_{i,j}=\pi (R_{c}^{2} -l_{i,j}^2/4)$ 
is the cross-sectional area between the  spheres of radius $R_c$ centered on the $i$-th and 
$j$-the node, respectively, $l_{i,j}$ is the distance between these centers, $\rho$ is the  
resistivity.
\par
By construction, each elemental resistance depends upon the distance between nodes.
Therefore, following up a conformational change, the variation of this distance 
implies a variation of each elemental resistance, which eventually leads to a variation
of the network resistance (and thus of the protein resistance) . 
As a consequence, a topological transformation can be monitored by means of resistance measurements.  
\par 
By taking the first and last amino acid of the protein primary structure as ideal electrical contacts, the protein resistance for a given $R_c$ is calculated by solving the corresponding linear resistance network within a standard Kirchhoff framework \cite{Nano}. 
Being interested to a change of the resistance rather than its absolute value, this choice for the contacts
does not modify the results of the model.
Numerical calculations show that for $R_c$ below a threshold value (typically around 4 \aa) 
the network is disconnected and thus not conducting at all.
By contrast, for $R_c$ well above the average distance between amino acids (typically above 15 \aa) \cite{Nano} each amino acid is connected with most of the others, thus the conduction becomes insensitive to modification of the structure.
Between these critical values of $R_c$, the change of resistance exhibits a smooth behaviour. 
Accordingly, we select the value
of $R_c$  which maximizes the change of resistance due to the conformational
change, and which is found to be around 6 \aa, a well accepted value for the interacting radius between amino acids  \cite{Tirion,Nano,Juanico}. 
\par  
To account for the strong super-linear \textit{I-V} characteristic, the model implements a barrier-limited current mechanism as follows.
At increasing voltage, each elemental resistance is allowed to take a second value of resistivity which, playing the role of a small series resistance of the network, is several order of magnitude lower than the first one. The probability of this choice mimics a barrier-limited mechanism in analogy with the case of an organic molecular layer \cite{Reed}. 
In this way, the initial linear increase of current with applied voltage turns into a superlinear increase, with the value of the barrier energy to be fitted by comparison with experiments.
Accordingly, the stochastic selection is taken to be ruled by the direct tunneling probability
\footnote{
We remark that the modelling leaves the possibility to replace tunneling by  a thermionic-emission mechanism, in case further experiments would evidence a significant temperature dependence of transport characteristics.}
\cite{Reed}:
\begin{equation} 
P_{i,j}= \exp[- \frac{2 l_{i,j}}{\hbar} \sqrt{2m(\Phi-eV_{i,j})} ]
\label{eq:1}
\end{equation}
where $V_{i,j}$ is the  potential drop between the  $i$-th and $j$-th node,
$m$ is an effective electron mass, here taken as that of the free electron, and $\Phi$ is the barrier height.
\par
We notice that the model does not take into account the protein environment
mainly for two reasons.
The former accounts for the fact that here we are interested only in the variation
of protein response, and this should be independent on the environment, since
it remains the same in the
conformational change. 
The latter reason is in agreement with ref.\cite{Jin1}, which concludes
that transmembrane electron transport occurs essentially only via bR and
not by the lipid bilayer.
Furthermore, the chromophore is  not included in the network being its contribution to the 
conductivity implicitly considered through the induced deformation of the structure. 
Otherwise, we do not exclude the possibility that the
chromophore can further contribute to the total conductivity.

\par
To make a reliable comparison between the native and activated
configuration, the corresponding representations
have to be taken from identical experimental settings \cite{Nano}.
Table 1 reports some characteristics of four couples of bR representations
which satisfy this requirement. Our model is strongly dependent on the protein
geometry, therefore, we focus our attention on the representations obtained
with the best resolution, (1M0K-1M0M) and (2NTU-2NTW). In the former couple
1M0K represent the K-state instead of the native state, and 1M0M refers to
the M-state. In the latter couple, 2NTU refers to the native state, while
2NTW refers to the L-state. Furthermore, we have verified that all the four
couples reported in table 1 show similar impedance change, \textit{i.e.}
the native state has a larger zero-frequency impedance with respect to the
activated state, any activation step is considered. In conclusion, we select
the couple (2NTU-2NTW) as computational sample because it has the best resolution,
contains the native state and although the activated state is that which
forthwith comes
before the M-state, we know that it can only underestimate our results, while
preserving their substantial validity.

\par 
Calculations proceed as follows.
The native and activated proteins are mapped into the corresponding topological networks.   
The resistance of the corresponding network at the given voltage is then calculated using the value 
$R_c$ = 6 \aa, which is found to give the best contrast. 
The value of $\rho$ for each elemental resistance is stochastically determined on the basis of the probability given in eq. (\ref{eq:1}) between the values $\rho_{MAX}$ and $\rho_{min}$. 
The value of $\rho_{MAX} = 10^{10} \ \Omega m$ is chosen to reproduce a current of about $10^{-19} \ A$ per bR monomer at 0.5 V, in accordance with experiments\cite{Jin}.
A rescaling of this value simply implies a rescaling of the total current \cite{Reed}.
Thus, since our interest focuses on the variation of current and not on its absolute value, the $\rho_{MAX}$  value serves as a normalizing parameter. 
The value of $\rho_{min} = 10^5 \ \Omega m$ is chosen to account for the superlinear behaviour of the \textit{I-V} characteristic and serves as series resistance of the circuit. 
In the voltage range considered in the experiments ($0 \div 1 \ V$) the deviation from linearity is sufficiently weak so that the value chosen for  $\rho_{min}$ does not play a significant role; indeed, the results do not change even  varying its value over 2 orders of magnitude. 
On the other hand, we found crucial  the role of the
barrier height, which drives the shape of the superlinear behaviour.
\par
The network resistance at the given voltage is then calculated by using the following  iterative procedure. 
First, the network is electrically solved by using the value $\rho_{MAX}$
for all the elemental resistances.
Second, each $\rho_{MAX}$ is stochastically replaced  by $\rho_{min}$  using the probability in eq. (2) according to the local potential drops calculated in the first step; the network is then electrically updated with the new \vspace{0.06cm}
distributed values of $\rho_{MAX,min}$. 
Third, the electrical 
update is iterated (typically $10^{4} \div 10^{5}$ iterations
depending on the value of the applied voltage) by repeating the second step until the resistance of the network, taken as the average value over the iteration steps, $<R>$,  converges within an uncertainty less than 1 \%. In this third step, the initial iterations ($100 \div 2000$ depending on voltage) contain a significant numerical noise and as such they are disregarded to avoid an unwanted drift of the average value (see fig. 1).
Finally, the current at the given voltage is calculated as $I=V/{<R>}$.
\par
The comparison between theory and experiments proceed as follows.
On the one hand, the theoretical model describes the \textit{I-V} modifications of a single protein when it undergoes a conformational change.
On the other hand, the experimental results are carried out on a macroscopic sample of 5 nm width and $2 \times 10^{-3} \ cm^2$ cross-sectional area.
Therefore, to compare the results, we normalize the  current value of the single protein in its native state at 1 V to that of the experiment in dark.
(The normalization corresponds to multiply the current of the single protein by a factor $10^{8}-10^{9}$).
The same normalization factor is used for the case of the single protein in its activated state, which is then compared
with the experimental value under green light.

\section{Results}
\begin{figure}
\onefigure[width=2.5in]{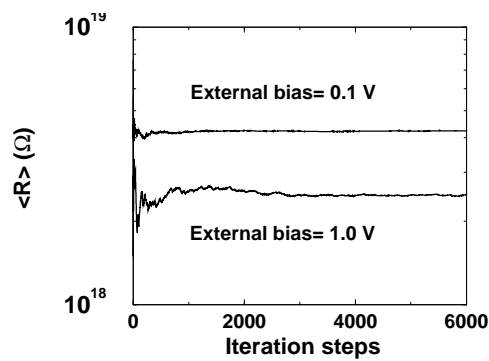}
 \caption{Running average of the network resistance for $\Phi$ = 59 meV. Data are obtained with an applied voltage of 0.1  and 1 V, respectively.}
\label{nanofig1}
 \end{figure}

Figure \ref{nanofig1} shows a typical evolution of the average resistance of the single protein at low and high voltages. 
Within the required uncertainty, at both voltages the resistance is found to converge above about $4 \times 10^3$ iteration steps, with an initial behaviour noisier at higher voltages. 
This evolution is typically observed for any value of $\Phi$  considered in calculations.
For the selected PDB entries reported in table 1, calculations give a resistance value of the native state larger than that of the activated state by about 10$~\%$ in the linear regime.
\par
\begin{figure}
 \onefigure[width=3.0in]{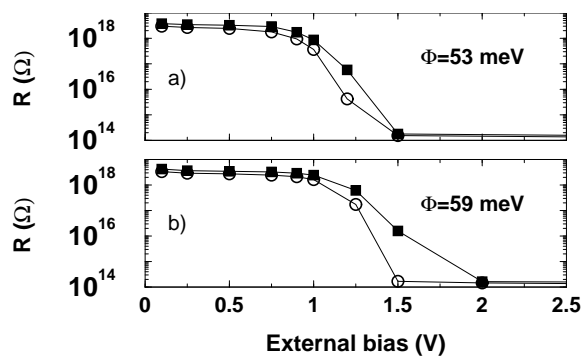}
 \caption{Network resistance as function of the applied voltage for the native (full squares) 
 and activated (open circles) state for the case of: (a) $\Phi$ = 53 meV
 and (b) $\Phi$ = 59 meV.
 Lines are guides to the eyes.}
 \label{nanofig2}
 \end{figure}

Figure \ref{nanofig2}  reports the single protein resistance \textit{vs}. the applied voltage for the native and activated state at two different values of the barrier height.
Here, we found that the sharp drop of the resistance starts at about 0.7 V, being more effective for 
$\Phi$ = 0.53 meV. 
Furthermore, in the region of voltages of interest for experiments, $0-1$ V, a further decrease of the series resistance is found to play a negligible role. 
\par
\begin{figure}
 \onefigure[width=3.0in]{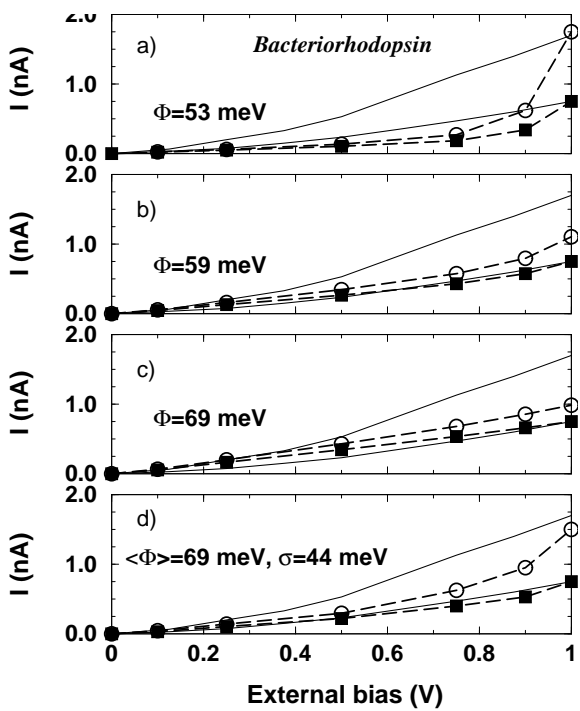}
 \caption{\textit{I-V} characteristics of the native and activated state of bR, 2NTU and 2NTW,
 respectively. 
 Full squares refer to the native state and open circles to activated state; dashed curves are guides to the eyes, 
 full curves  refer to experiments \cite{Jin}.
 Calculations refer to  different barrier heights: (a) $\Phi=53$ meV, (b)  $\Phi=59$ meV, (c) $\Phi=69$ meV,
 (d) average  $\Phi=69$ meV Gaussian distributed with a variance $\sigma$ = 44 meV.
 All the currents obtained by calculations are normalized to the experimental
 value of current in the dark at 1 V \cite{Jin}.}
 \label{nanofig3}
 \end{figure}

 
Figures \ref{nanofig3} report a quantitative comparison between theory and experiments. 
The experimental results, as detailed in ref. \cite{Jin1},
give the net message that green light illumination induces an increasing
of current in bR junctions. 
The value of this increase depends on
the junction preparation (bR/PC vescicles or bR/OTG vescicles), and has a
significant range of variability, from a minimum of about 30$\%$ to a maximum around $300\%$.
Accordingly, taking into account this experimental uncertainty, the comparison between theory and experiments is mostly qualitative. Overall, we are satisfied that the global behaviour, like the increase of current with illumination, and the shape of experimental curves, is well reproduced.
In details, figs. 3(a), 3(b) and 3(c) report the comparison of the \textit{I-V} characteristics of bR/PC 
junctions \cite{Jin}, obtained by a fine tuning of the barrier height with, respectively, $\Phi$= 53, 59, 69 meV.
For $\Phi$= 59 meV (Fig. 3(b))  calculations well reproduce the \textit{I-V} characteristic of the native
state, but underestimate that of the activated state. 
On the other hand, a further increase to $\Phi$ = 69 meV does not fully account for the superlinear behaviour
(see fig. 3 (c)).
Therefore, in analogy with the case of disordered organic materials \cite{Baranovski,Bassler},  fig. 3 (d) reports the results obtained by taking a Gaussian distribution of $\Phi$ within an average value of 69 meV and a dispersion $\sigma$ = 44 meV.
We conclude that the results reported in fig. 3 (in particular fig. 3 (d)) capture the essence of the experiments and validate the conjecture that {\it the conformational change of bacteriorhodopsin can be 
reliably detected by current transport measurements in the presence and absence of light}. 
\par
Remarkably, we notice that,  when scaled by a factor of 25, the values used
here for the electron effective mass and barrier heights are in reasonable
agreement with the values obtained within a simple direct tunneling model of charge transport like in a metal-insulator-metal structure\cite{Gomila}.
Indeed, the factor 25 represents the mean number of sequential steps made by a carrier when going from one contact to the opposite one. Clearly, the advantage of the present model stems from the strict correlation between the microscopic structure of the protein and its macroscopic electrical property.
\par
By construction, the present model reproduces also other details of experiments
summarized in the introduction.
\par
Theory predicts that in absence of a conformational change the protein resistance
remains the same. This is verified experimentally in several ways: i) By substituting the original chromophore with one that is not sensible to the light, and finding that the \textit{I-V} sensitivity to the green light is washed out. ii) By substituting the retinal with indeed with retinaloxime \cite{Jin,Jin1}.
Then green light does not
 modify the conformation and thus does not produce variation in the (nonlinear)
 current. iii) By using an artificial pigment derived from "locked" retinal. In this case the protein reaches an M state different
 from the wild M state and there is only a small current increase \cite{Jin1}.

\par
Theory predicts the presence or absence of the protein conductance on the basis of the range of interaction 
between amino acids. Absence of conduction is achieved by taking a small interaction radius (typically less
than 3.8 \aa).
This is verified by the experiments carried out on proteins deprived of the retinal, where the response 
to an external bias is found to be a very low and completely noisy signal \cite{Jin, Jin1}. 
The origin of this result is interpreted as the protein loss of connectivity \textit{i.e.} 
as a dramatic deterioration of its structure.

\par
Theory predicts the symmetry of the \textit{I-V} characteristic in sweeping from 
negative-to-positive bias, which
is in accordance with experimental results. 
\par
To further test the physical plausibility of the model presented here, 
we have considered the case of (bovine) rhodopsin, a photoreceptor
pertaining to the G-protein-coupled receptor (GPCR) family. 
Although the mechanism of photo-activation is essentially different from that of bR, this protein
shares with bR a quite similar tertiary structure, \textit{i.e.} both are
seven-helices transmembrane
proteins. 
This fact induced the researchers to take the bR structure as a
template for reconstructing the 3D conformation of rhodopsin (and more generally of all
the GPCR proteins). 
Accordingly, calculations have been carried out for the native and activated state
of rhodopsin, whose structures were engineered as reported in \cite{Nano}.
The obtained results, which parallel those of fig. 3(b), are reported in fig. 4.
Remarkably, for the analogous conformational change, the model predicts that rhodopsin exhibits
a  behaviour opposite to that of bR, \textit{i.e.} in going from the natural to the activated configuration, current
at a given voltage is suppressed.   
Physically, we believe that this opposite behaviour follows from the opposite change the retinal undergoes
when absorbing a photon. 
In rhodopsin the retinal shape  goes from bent to straight, while 
the reverse occurs in bR \cite{PDB}. 
As a consequence, we arrive at the plausible conclusion that an opposite conformational change of the entire protein should imply an opposite change of its \textit{I-V} characteristics. 
\section{Conclusions}
We have presented a resistor network model implemented for a sequential tunneling mechanism to investigate charge transport, and in particular its structure dependence,  in proteins. 
The model proves its capability  to capture the correlation between a change of the protein conformation  and that of its current-voltage characteristic.
The theory is validated on recent experiments performed on a nanojunction filled with bacteriorhodopsin, a light-activated protein, which evidences a substantial increase of the current, at a given voltage, in going from dark to the presence of a green light. 
A good agreement with experimental data is obtained  when considering  an interaction radius between the amino acids of $6$ \aa  and an average barrier energy of 69 meV with a Gaussian distribution of standard deviation $\sigma$ = 44 meV.
Interestingly, the model is able to explain other details of experiments, such as the role played by the presence or absence of the chromophore responsible of the sensing activity, as well as it is compatible with simple direct tunnelling models developed on the metal-insulator-metal scheme.     
Under analogous conditions, application of the model to the case of bovine rhodopsin  predicts an
opposite behaviour with a suppression of the current. 
Confirmation of such a prediction remains an experimental challenge.   
  \begin{figure}
 \onefigure[width=3.0in]{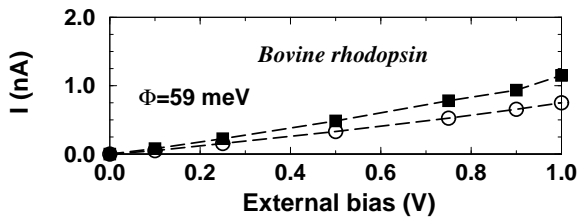}
 \caption{Predicted \textit{I-V} characteristics of the native (full squares) and activated (open circles) state of bovine rhodopsin,
 dashed curves are guides to the eyes. 
  Calculations are carried out analogously to those in fig. 3(b). All the currents obtained by calculations are normalized to the experimental
 value of current in the dark at 1 V \cite{Jin}.}
 \label{nanofig4}
 \end{figure} 
\begin{largetable}
\caption{Bacteriorhodopsin PDB entries}
\label{tab.1}
\begin{tabular}{|clccc|}
\hline
Entry  & State & Resolution(\aa)& Temperature (K)& Amino acids\\
\hline\hline
1FBB& Native & 3.20 & -- &224\\
1FBK& Activated & 3.20 & -- &225\\
\hline
1QM8& Native & 2.50 & 100 &229\\
1DZE& M State & 2.50 & 100 &224\\
\hline
2NTU& Native & 1.53 & 100 &222\\
2NTW& L State & 1.53 & 100 &222\\
\hline
1M0K& K State & 1.43 & 100 &222\\
1M0M& M State & 1.43 & 100 &224\\
\hline
\end{tabular}
\end{largetable}
%


\acknowledgments
The authors acknowledge Drs C. Pennetta and G. Gomila for valuable discussions on the subject.
This research is supported by EC through the Bioelectronic Olfactory Neuron Device (BOND) project.





\end{document}